\documentclass[twocolumn,aps,tightenlinefs,superscriptaddress,preprintnumbers]{revtex4}
\usepackage{graphicx,color,dcolumn,booktabs,bm}
\usepackage{longtable,lscape}
\usepackage{amsmath}
\usepackage{indentfirst}
\usepackage{epsfig}
\usepackage{feynmf}
\usepackage{epstopdf}
\usepackage{slashed}
\usepackage{cases}
\usepackage{color}
\usepackage{multirow}
\usepackage{ulem}
\usepackage{graphicx,color,dcolumn,booktabs,bm}
\usepackage{verbatim}
\usepackage[colorlinks,linkcolor=black,anchorcolor=green,citecolor=blue]{hyperref}
\usepackage{mathrsfs}
\usepackage{rotating}
\usepackage{threeparttable}
\usepackage{dcolumn}

\newcommand{\eff}{\varepsilon}
\newcommand{\BR}{{\cal B}}

\newcommand{\pip}{\pi^+}
\newcommand{\pim}{\pi^-}

\newcommand{\EE}{e^+e^-}

\newcommand{\beq}{\begin{equation}}
\newcommand{\eeq}{\end{equation}}
\newcommand{\bitm}{\begin{itemize}}
\newcommand{\eitm}{\end{itemize}}

\newcommand{\ones}{\Upsilon(1S)}
\newcommand{\twos}{\Upsilon(2S)}

\newcommand{\threes}{\Upsilon(3S)}
\newcommand{\fours}{\Upsilon(4S)}
\newcommand{\fives}{\Upsilon(5S)}

\begin{document}
\hyphenpenalty=10000


\title{\quad\\[0.1cm]\boldmath Measurements of the branching fractions of $\Xi_c^0 \to \Lambda K_S^0$, $\Xi_c^0 \to \Sigma^0 K_S^0$,
and $\Xi_c^0 \to \Sigma^+ K^-$ decays at Belle}

\noaffiliation
\affiliation{Department of Physics, University of the Basque Country UPV/EHU, 48080 Bilbao}
\affiliation{University of Bonn, 53115 Bonn}
\affiliation{Brookhaven National Laboratory, Upton, New York 11973}
\affiliation{Budker Institute of Nuclear Physics SB RAS, Novosibirsk 630090}
\affiliation{Faculty of Mathematics and Physics, Charles University, 121 16 Prague}
\affiliation{Chonnam National University, Gwangju 61186}
\affiliation{Chung-Ang University, Seoul 06974}
\affiliation{University of Cincinnati, Cincinnati, Ohio 45221}
\affiliation{Deutsches Elektronen--Synchrotron, 22607 Hamburg}
\affiliation{Duke University, Durham, North Carolina 27708}
\affiliation{Institute of Theoretical and Applied Research (ITAR), Duy Tan University, Hanoi 100000}
\affiliation{University of Florida, Gainesville, Florida 32611}
\affiliation{Department of Physics, Fu Jen Catholic University, Taipei 24205}
\affiliation{Key Laboratory of Nuclear Physics and Ion-beam Application (MOE) and Institute of Modern Physics, Fudan University, Shanghai 200443}
\affiliation{Justus-Liebig-Universit\"at Gie\ss{}en, 35392 Gie\ss{}en}
\affiliation{Gifu University, Gifu 501-1193}
\affiliation{II. Physikalisches Institut, Georg-August-Universit\"at G\"ottingen, 37073 G\"ottingen}
\affiliation{SOKENDAI (The Graduate University for Advanced Studies), Hayama 240-0193}
\affiliation{Gyeongsang National University, Jinju 52828}
\affiliation{Department of Physics and Institute of Natural Sciences, Hanyang University, Seoul 04763}
\affiliation{University of Hawaii, Honolulu, Hawaii 96822}
\affiliation{High Energy Accelerator Research Organization (KEK), Tsukuba 305-0801}
\affiliation{J-PARC Branch, KEK Theory Center, High Energy Accelerator Research Organization (KEK), Tsukuba 305-0801}
\affiliation{National Research University Higher School of Economics, Moscow 101000}
\affiliation{Forschungszentrum J\"{u}lich, 52425 J\"{u}lich}
\affiliation{IKERBASQUE, Basque Foundation for Science, 48013 Bilbao}
\affiliation{Indian Institute of Science Education and Research Mohali, SAS Nagar, 140306}
\affiliation{Indian Institute of Technology Bhubaneswar, Satya Nagar 751007}
\affiliation{Indian Institute of Technology Guwahati, Assam 781039}
\affiliation{Indian Institute of Technology Hyderabad, Telangana 502285}
\affiliation{Indian Institute of Technology Madras, Chennai 600036}
\affiliation{Indiana University, Bloomington, Indiana 47408}
\affiliation{Institute of High Energy Physics, Chinese Academy of Sciences, Beijing 100049}
\affiliation{Institute of High Energy Physics, Vienna 1050}
\affiliation{Institute for High Energy Physics, Protvino 142281}
\affiliation{INFN - Sezione di Napoli, I-80126 Napoli}
\affiliation{INFN - Sezione di Roma Tre, I-00146 Roma}
\affiliation{INFN - Sezione di Torino, I-10125 Torino}
\affiliation{Iowa State University, Ames, Iowa 50011}
\affiliation{Advanced Science Research Center, Japan Atomic Energy Agency, Naka 319-1195}
\affiliation{J. Stefan Institute, 1000 Ljubljana}
\affiliation{Institut f\"ur Experimentelle Teilchenphysik, Karlsruher Institut f\"ur Technologie, 76131 Karlsruhe}
\affiliation{Kavli Institute for the Physics and Mathematics of the Universe (WPI), University of Tokyo, Kashiwa 277-8583}
\affiliation{Department of Physics, Faculty of Science, King Abdulaziz University, Jeddah 21589}
\affiliation{Kitasato University, Sagamihara 252-0373}
\affiliation{Korea Institute of Science and Technology Information, Daejeon 34141}
\affiliation{Korea University, Seoul 02841}
\affiliation{Kyoto Sangyo University, Kyoto 603-8555}
\affiliation{Kyungpook National University, Daegu 41566}
\affiliation{P.N. Lebedev Physical Institute of the Russian Academy of Sciences, Moscow 119991}
\affiliation{Liaoning Normal University, Dalian 116029}
\affiliation{Faculty of Mathematics and Physics, University of Ljubljana, 1000 Ljubljana}
\affiliation{Ludwig Maximilians University, 80539 Munich}
\affiliation{Luther College, Decorah, Iowa 52101}
\affiliation{Malaviya National Institute of Technology Jaipur, Jaipur 302017}
\affiliation{Faculty of Chemistry and Chemical Engineering, University of Maribor, 2000 Maribor}
\affiliation{Max-Planck-Institut f\"ur Physik, 80805 M\"unchen}
\affiliation{School of Physics, University of Melbourne, Victoria 3010}
\affiliation{University of Mississippi, University, Mississippi 38677}
\affiliation{University of Miyazaki, Miyazaki 889-2192}
\affiliation{Moscow Physical Engineering Institute, Moscow 115409}
\affiliation{Graduate School of Science, Nagoya University, Nagoya 464-8602}
\affiliation{Kobayashi-Maskawa Institute, Nagoya University, Nagoya 464-8602}
\affiliation{Universit\`{a} di Napoli Federico II, I-80126 Napoli}
\affiliation{Nara Women's University, Nara 630-8506}
\affiliation{National Central University, Chung-li 32054}
\affiliation{Department of Physics, National Taiwan University, Taipei 10617}
\affiliation{H. Niewodniczanski Institute of Nuclear Physics, Krakow 31-342}
\affiliation{Nippon Dental University, Niigata 951-8580}
\affiliation{Niigata University, Niigata 950-2181}
\affiliation{University of Nova Gorica, 5000 Nova Gorica}
\affiliation{Novosibirsk State University, Novosibirsk 630090}
\affiliation{Osaka City University, Osaka 558-8585}
\affiliation{Pacific Northwest National Laboratory, Richland, Washington 99352}
\affiliation{Panjab University, Chandigarh 160014}
\affiliation{University of Pittsburgh, Pittsburgh, Pennsylvania 15260}
\affiliation{Punjab Agricultural University, Ludhiana 141004}
\affiliation{Research Center for Nuclear Physics, Osaka University, Osaka 567-0047}
\affiliation{Meson Science Laboratory, Cluster for Pioneering Research, RIKEN, Saitama 351-0198}
\affiliation{Dipartimento di Matematica e Fisica, Universit\`{a} di Roma Tre, I-00146 Roma}
\affiliation{Department of Modern Physics and State Key Laboratory of Particle Detection and Electronics, University of Science and Technology of China, Hefei 230026}
\affiliation{Showa Pharmaceutical University, Tokyo 194-8543}
\affiliation{Soongsil University, Seoul 06978}
\affiliation{Sungkyunkwan University, Suwon 16419}
\affiliation{School of Physics, University of Sydney, New South Wales 2006}
\affiliation{Department of Physics, Faculty of Science, University of Tabuk, Tabuk 71451}
\affiliation{Tata Institute of Fundamental Research, Mumbai 400005}
\affiliation{Department of Physics, Technische Universit\"at M\"unchen, 85748 Garching}
\affiliation{School of Physics and Astronomy, Tel Aviv University, Tel Aviv 69978}
\affiliation{Toho University, Funabashi 274-8510}
\affiliation{Department of Physics, Tohoku University, Sendai 980-8578}
\affiliation{Earthquake Research Institute, University of Tokyo, Tokyo 113-0032}
\affiliation{Department of Physics, University of Tokyo, Tokyo 113-0033}
\affiliation{Tokyo Institute of Technology, Tokyo 152-8550}
\affiliation{Tokyo Metropolitan University, Tokyo 192-0397}
\affiliation{Virginia Polytechnic Institute and State University, Blacksburg, Virginia 24061}
\affiliation{Wayne State University, Detroit, Michigan 48202}
\affiliation{Yamagata University, Yamagata 990-8560}
\affiliation{Yonsei University, Seoul 03722}
\author{Y.~Li}\affiliation{Key Laboratory of Nuclear Physics and Ion-beam Application (MOE) and Institute of Modern Physics, Fudan University, Shanghai 200443} 
\author{J.~X.~Cui}\affiliation{Key Laboratory of Nuclear Physics and Ion-beam Application (MOE) and Institute of Modern Physics, Fudan University, Shanghai 200443} 
\author{S.~Jia}\affiliation{Key Laboratory of Nuclear Physics and Ion-beam Application (MOE) and Institute of Modern Physics, Fudan University, Shanghai 200443} 
\author{C.~P.~Shen}\affiliation{Key Laboratory of Nuclear Physics and Ion-beam Application (MOE) and Institute of Modern Physics, Fudan University, Shanghai 200443} 
\author{I.~Adachi}\affiliation{High Energy Accelerator Research Organization (KEK), Tsukuba 305-0801}\affiliation{SOKENDAI (The Graduate University for Advanced Studies), Hayama 240-0193} 
\author{J.~K.~Ahn}\affiliation{Korea University, Seoul 02841} 
\author{H.~Aihara}\affiliation{Department of Physics, University of Tokyo, Tokyo 113-0033} 
\author{S.~Al~Said}\affiliation{Department of Physics, Faculty of Science, University of Tabuk, Tabuk 71451}\affiliation{Department of Physics, Faculty of Science, King Abdulaziz University, Jeddah 21589} 
\author{D.~M.~Asner}\affiliation{Brookhaven National Laboratory, Upton, New York 11973} 
\author{H.~Atmacan}\affiliation{University of Cincinnati, Cincinnati, Ohio 45221} 
% ...
\collaboration{The Belle Collaboration}

\begin{abstract}
Using the entire data sample of $980\mathrm{~fb}^{-1}$ collected with the Belle detector at the KEKB asymmetric-energy $e^+e^-$ collider,
we present measurements of the branching fractions of the Cabibbo-favored decays $\Xi_c^0 \to \Lambda K_S^0$,
$\Xi_c^0 \to \Sigma^0 K_S^0$, and $\Xi_c^0 \to \Sigma^+ K^-$. Taking the decay $\Xi_c^0 \to \Xi^- \pip$ as the normalization mode,
we measure the branching fraction ratio $\BR(\Xi_c^0 \to \Lambda K_S^0)/\BR(\Xi_c^0 \to \Xi^- \pip)
= 0.229\pm0.008\pm0.012$ with improved precision, and measure the branching fraction ratios
$\BR(\Xi_c^0 \to \Sigma^0 K_S^0)/\BR(\Xi_c^0 \to \Xi^- \pip) = 0.038\pm0.006\pm0.004$ and $\BR(\Xi_c^0 \to \Sigma^+ K^-)/\BR(\Xi_c^0 \to \Xi^- \pip) = 0.123\pm0.007\pm0.010$ for the first time. Taking into account the branching fraction of the normalization mode,
the absolute branching fractions are determined to be
$\BR(\Xi_c^0 \to \Lambda K_S^0) = (3.27\pm0.11\pm0.17\pm0.73) \times 10^{-3}$,
$\BR(\Xi_c^0 \to \Sigma^0 K_S^0) = (0.54\pm 0.09\pm 0.06\pm 0.12) \times 10^{-3}$, and
$\BR(\Xi_c^0 \to \Sigma^+ K^-) = (1.76\pm 0.10\pm0.14\pm 0.39) \times 10^{-3}$. The first
and second uncertainties above are statistical and systematic, respectively,
while the third ones arise from the uncertainty of the branching fraction of $\Xi_c^0 \to \Xi^- \pip$.
\end{abstract}

\maketitle
\section{\boldmath Introduction}
Charmed baryons provide a unique laboratory to study the subtle interplay
of the strong and weak interactions. Recently, there have been several major breakthroughs
in the experimental study of the $\Xi_{c}^{0}$ baryon. Belle has presented the first
measurement of the absolute branching fraction $\BR(\Xi_c^0 \to \Xi^- \pi^+) = (1.80 \pm 0.50 \pm 0.14)\%$~\cite{Abs_Xic0},
so that the branching fractions of other decay channels of $\Xi_c^0$ can be determined from ratios
of branching fractions. The branching fractions of the semileptonic
decays $\Xi_c^0 \to \Xi^- e^+ \nu_e$ and $\Xi_c^0 \to \Xi^- \mu^+ \nu_\mu$ have been
measured to be $(1.31 \pm 0.04 \pm 0.07 \pm 0.38)$\% and
$(1.27 \pm 0.06 \pm 0.10 \pm 0.37)$\%~\cite{semi_Xic},
where the uncertainties are statistical, systematic, and
from $\BR(\Xi_c^0 \to \Xi^- \pi^+)$, respectively. The corresponding branching fraction
ratio $\BR(\Xi_c^0 \to \Xi^- e^+ \nu_e)/\BR(\Xi_c^0 \to \Xi^- \mu^+ \nu_\mu)$ is
$1.03\pm0.09$,  which is consistent with the expectation
of lepton flavor universality. Very recently, the branching fractions and asymmetry
parameters of the Cabibbo-favored (CF) decays $\Xi_c^0 \to \Lambda\bar{K}^{*0}$, $\Xi_c^0 \to \Sigma^0\bar{K}^{*0}$,
and $\Xi_c^0 \to \Sigma^+ K^{*-}$ have been measured for the first time~\cite{Xic_jia}.

\begin{figure}[htbp]
	\begin{center}
		\includegraphics[width=4.25cm]{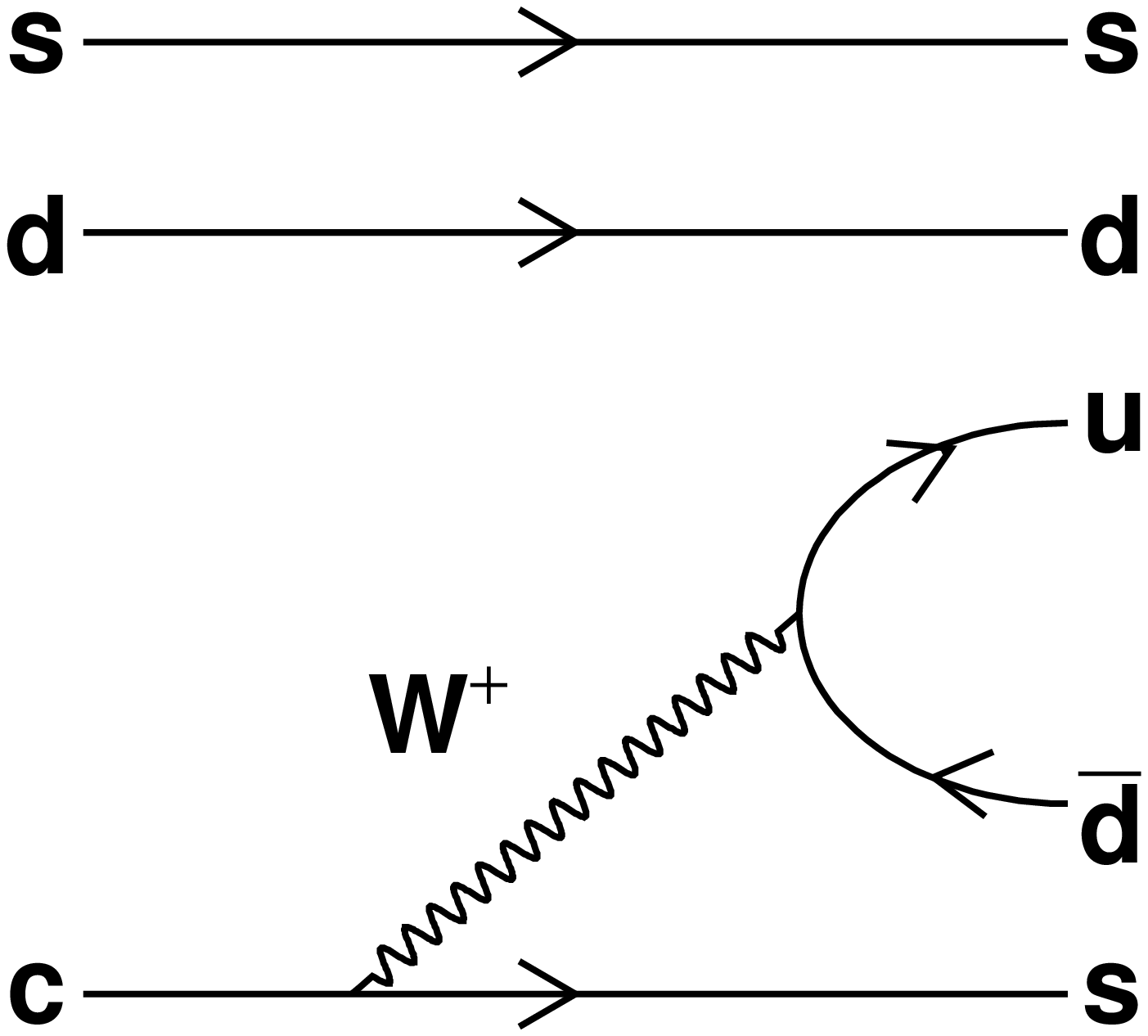}
		\includegraphics[width=4.23cm]{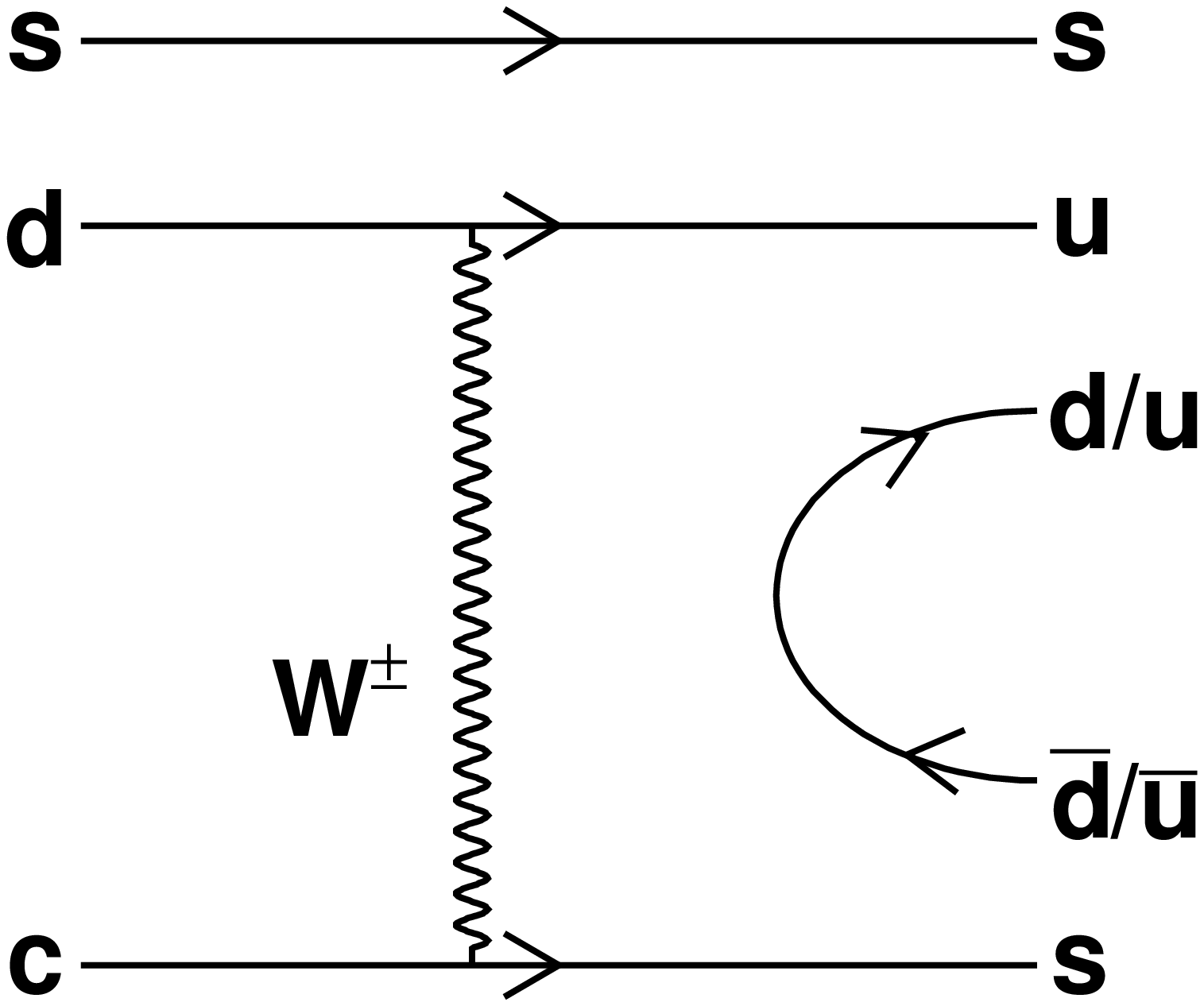}
		\put(-240,92){\bf (a)} \put(-125,92){\bf (b)}
		\caption{Feynman diagrams from (a) internal $W$-emission for $\Xi_c^0 \to \Lambda \bar{K}^0/\Sigma^0\bar{K}^0$
			 decays and (b) $W$-exchange for $\Xi_c^0 \to \Lambda \bar{K}^0/\Sigma^0\bar{K}^0/\Sigma^+K^-$ decays.}\label{fig1}
	\end{center}
\end{figure}

Theoretical calculations for the two-body hadronic weak decays $\Xi_c^0 \to B + P$
have been performed using dynamical models~\cite{Zou:2019kzq} and $SU(3)_F$ flavor symmetry
methods~\cite{Geng:2019xbo, Zhao:2018mov}, where $B$ and $P$ represent light baryons and
pseudoscalar mesons. In hadronic weak decays of charmed baryons, nonfactorizable contributions from inner
$W$-emission and $W$-exchange topological diagrams play an essential role and cannot be neglected,
in contrast with their negligible effects in heavy meson decays~\cite{Cheng:2021qpd}.
Figure~\ref{fig1} shows the Feynman diagrams from internal $W$-emission
for $\Xi_c^0 \to \Lambda \bar{K}^0/\Sigma^0\bar{K}^0$ decays and $W$-exchange for $\Xi_c^0 \to \Lambda \bar{K}^0/\Sigma^0\bar{K}^0/\Sigma^+K^-$ decays as examples.
In Ref.~\cite{Zou:2019kzq}, the authors found that the factorizable
and nonfactorizable terms in both the $S$- and $P$-wave amplitudes of the decay $\Xi_c^0 \to \Sigma^0 \bar{K}^0$ interfere destructively,
resulting in a small branching fraction. On the other hand, the interference in the decay
$\Xi_c^0 \to \Lambda \bar{K}^0$ is found to be constructive. The decay $\Xi_c^0 \to \Sigma^+ K^-$
proceeds only through purely nonfactorizable diagrams, and it allows us to check the importance of such decay diagrams.
The branching fractions of $\Xi_c^0 \to \Lambda \bar{K}^0$,
$\Xi_c^0 \to \Sigma^0 \bar{K}^0$, and $\Xi_c^0 \to \Sigma^+ K^-$ decays predicted by different
theoretical models are listed in Table~\ref{tab:predicted}.

\begin{table}[htbp]
	\caption{\label{tab:predicted} The predicted branching fractions in units of $10^{-3}$ for
	the CF decays $\Xi_c^0 \to \Lambda \bar{K}^0/\Sigma^0 \bar{K}^0/\Sigma^+ K^- $ based on
	dynamical model calculations and $SU(3)_{F}$ flavor symmetry approaches.}
	\begin{tabular}{lccc}
		\hline\hline
		Modes & Zou {\it et al.}~\cite{Zou:2019kzq} & Geng {\it et al.}~\cite{Geng:2019xbo} & Zhao {\it et al.}~\cite{Zhao:2018mov}  \\
		\hline
		$\Xi_c^0 \to \Lambda  \bar{K}^0$  & $13.3$ & $10.5 \pm 0.6$  & $8.3 \pm 5.0$  \\
		$\Xi_c^0 \to \Sigma^0 \bar{K}^0$  &  $0.4$ & $ 0.8 \pm 0.8$  & $7.9 \pm 4.8$ \\	
		$\Xi_c^0 \to \Sigma^+      K^- $  &  $7.8$ & $ 5.9 \pm 1.1$  & $22.0\pm 5.7$  \\	
		\hline\hline
	\end{tabular}
\end{table}

The ratio of the branching fraction of $\Xi_c^0 \to \Lambda K_S^0$ relative to that of
$\Xi_c^0 \to \Xi^- \pip$ has been measured to be $0.21 \pm 0.02 \pm 0.02$
by Belle using a $140\mathrm{~fb}^{-1}$ data sample~\cite{Belle:2004tfm}.
In this paper, we measure the branching fraction ratio
$\BR(\Xi_c^0 \to \Lambda K_S^0)/\BR(\Xi_c^0 \to \Xi^- \pip)$
to improve the precision, and present the first measurements of the branching fraction ratios
$\BR(\Xi_c^0 \to \Sigma^0 K_S^0)/\BR(\Xi_c^0 \to \Xi^- \pip)$ and
$\BR(\Xi_c^0 \to \Sigma^+ K^-)/\BR(\Xi_c^0 \to \Xi^- \pip)$ using the entire data sample
of $980\mathrm{~fb}^{-1}$ collected with the Belle detector. Charge-conjugate modes are also implied unless
otherwise stated throughout this paper.

\section{\boldmath The data sample and the belle detector}
This analysis is based on data recorded at or near the $\ones$, $\twos$, $\threes$,
$\fours$, and $\fives$ resonances by the Belle detector~\cite{detector1, detector2}
at the KEKB \mbox{asymmetric-energy} $e^+e^-$ collider~\cite{collider1, collider2}.
The total data sample corresponds to an integrated luminosity of $980\mathrm{~fb}^{-1}$~\cite{detector2}.
The detector is
described in detail elsewhere~\cite{detector1, detector2}.

Monte Carlo (MC) simulated signal events are generated using {\sc EvtGen}~\cite{evtgen}
to optimize the signal selection criteria and calculate the reconstruction efficiencies.
Events for the $\EE \to c\bar{c}$ production are generated using {\sc PYTHIA}~\cite{pythia} with a
specific Belle configuration, where one of the two charm quarks hadronizes into a $\Xi_{c}^0$ baryon.
The $\Xi_c^0 \to \Lambda K_S^0/\Sigma^0 K_S^0/\Sigma^+ K^-$ decays are generated using a phase
space model. The simulated events are processed with a detector simulation based
on {\sc GEANT3}~\cite{geant}. Inclusive MC samples of $\Upsilon(1S,2S,3S)$ decays,
$\fours \to B^{+}B^{-}/B^{0}\bar{B}^{0}$,  $\fives \to B_{(s)}^{(*)} \bar{B}_{(s)}^{(*)}$,
and $\EE \to q\bar{q}$ ($q=u,\,d,\,s,\,c$) at center-of-mass (C.M.) energies of $9.460$, $10.024$,
$10.355$, $10.520$, $10.580$, and $10.867\mathrm{~GeV}$ corresponding to the total integrated luminosity of data are used to check possible
peaking backgrounds and to verify the event selection criteria.

\section{\boldmath Common Event selection criteria}

The selection of the photon candidates as well as the particle identifications (PID) of kaon, pion, and proton
are performed using the same methods as in Ref.~\cite{Xic_jia}.
Furthermore, the impact parameters
of kaons with respect to the interaction point (IP) are required to be less than 0.2~cm
and $1.0\mathrm{~cm}$ perpendicular to, and along the beam direction, respectively.

The $K_{S}^{0}$ candidates are first reconstructed from pairs of oppositely charged tracks,
which are treated as pions, with a production vertex significantly separated from the
IP, and then selected using an artificial neural network~\cite{neural,input}.
The $\Lambda$ candidates are reconstructed via $\Lambda \to p\pi^-$ decays.
The invariant masses of the
$K_S^0$ and $\Lambda$ candidates are required to be within $9.5\mathrm{~MeV}/c^2$ and
$3.5\mathrm{~MeV}/c^2$  of the corresponding nominal masses~\cite{PDG} ($>95\%$ signal events are retained), respectively.


For the $\Sigma^0 \to \Lambda \gamma$ reconstruction, the selected $\Lambda$ candidate is
combined with a photon to form a $\Sigma^0$ candidate. The energy of the photon is required
to exceed $130\mathrm{~MeV}$ in the laboratory frame to suppress combinatorial backgrounds.
This criterion is optimized by maximizing the figure-of-merit $N_{\rm sig}/\sqrt{N_{\rm sig} + N_{\rm bkg}}$,
where $N_{\rm sig}$ is the number of expected signal events of $\Xi_c^0 \to \Sigma^0 K_S^0$ decay,
and $N_{\rm bkg}$ is the number of background events
in the normalized $\Xi_c^0$ sidebands in data. $N_{\rm sig}$ is obtained from the following formula

\begin{align*}
N_{\rm sig} =~& \eff(\Xi_c^0 \to \Sigma^0 K_S^0) \times \frac{N^{\rm obs}(\Xi_{c}^0 \to \Xi^- \pip)}{\eff(\Xi_c^0 \to \Xi^- \pip)} \\ & \times \frac{\BR(\Xi_{c}^0 \to \Sigma^0 K_S^0) \BR(\Sigma^0 \to \Lambda \gamma) \BR(K_S^0 \to \pip \pim)}{\BR(\Xi_c^0 \to \Xi^- \pip)\BR(\Xi^- \to \Lambda \pim)},
\end{align*}
where $\eff(\Xi_c^0 \to \Sigma^0 K_S^0)$ and $\eff(\Xi_c^0 \to \Xi^- \pip)$
are the reconstruction efficiencies of $\Xi_c^0 \to \Sigma^0 K_S^0$ and
$\Xi_c^0 \to \Xi^- \pip$ decays; $N^{\rm obs}(\Xi_{c}^0 \to \Xi^- \pip)$
is the number of observed $\Xi_c^0 \to \Xi^- \pip$ signal events in data;
$\BR(\Xi_{c}^0 \to \Sigma^0 K_S^0) = 2.0\times10^{-4}$ is the branching
fraction of $\Xi_c^0 \to \Sigma^0 K_S^0$ decay predicted by dynamical model
calculations~\cite{Zou:2019kzq},  $\BR(\Sigma^0 \to \Lambda \gamma) = 100\%$,
$\BR(K_S^0 \to \pip \pim)$ = $(69.20 \pm 0.05)\%$, $\BR(\Xi_{c}^0 \to \Xi^- \pip) =
(1.43 \pm 0.32)\%$~\cite{PDG}, and $\BR(\Xi^- \to \Lambda \pim) = (99.887 \pm 0.035)\%$~\cite{PDG}.
The optimized selection criterion is the same using the assumed branching
fractions from the above-mentioned theoretical predictions.

The $\Sigma^+\to p \pi^0$ reconstruction is performed as follows~\cite{sigma}.
Photon pairs are kept as $\pi^0$ candidates. The reconstructed invariant mass
of the $\pi^0$ candidates is required to be within $15\mathrm{~MeV}/c^2$ of the $\pi^0$ nominal mass~\cite{PDG},
corresponding to approximately twice the resolution. To reduce the combinatorial backgrounds,
the momentum of the $\pi^0$ in the $\EE$ C.M.\ frame is required to exceed $0.3\mathrm{~GeV}/c$, which is
optimized using the same method that was used for the energy of photon from the $\Sigma^0$ decay~\cite{input_Br}.
Combinations of $\pi^0$ candidates and protons are made using those protons with a significantly
large ($> 1\mathrm{~mm}$) distance of closest approach to the IP. Then, taking the IP as the point
of origin of the $\Sigma^+$, the sum of the proton and $\pi^0$ momenta is taken as the momentum
vector of the $\Sigma^+$ candidate. The intersection of this trajectory with the reconstructed
proton trajectory is then found and this position is taken as the decay location of the
$\Sigma^+$ baryon. The $\pi^0$ is then refit using this location as its point of origin.
Only those combinations with the decay location of the $\Sigma^+$ indicating a positive
$\Sigma^+$ pathlength are retained.

The $\Lambda K_S^0$, $\Sigma^0 K_S^0$, or $\Sigma^+ K^-$ combinations are made to form $\Xi_c^0$ candidates
with their daughter tracks fitted to a common vertex. The helicity angle of $\Xi_c^0$ candidates is required to be $\lvert\mathrm{cos}\theta(\Xi^0_c)\rvert < 0.75$
to suppress the combinatorial background, where $\theta(\Xi_c^0)$ is the angle between the $\Lambda/\Sigma^0/\Sigma^+$ momentum vector and the boost direction from the laboratory frame in the $\Xi_c^0$ rest frame. To reduce combinatorial backgrounds, especially from \mbox{$B$-meson} decays, the scaled momentum $x_{p} = p^{*}_{\Xi_{c}^0}$/$p_{\rm max}$ is required to be larger than $0.55$. Here, $p^{*}_{\Xi_{c}^0}$ is the momentum of $\Xi_{c}^0$ candidates in the $\EE$ C.M.\ frame, and $p_{\rm max}=\frac{1}{c}\sqrt{E^2_{\rm beam}-M_{\Xi_c^0}^2 c^4}$, where $E_{\rm beam}$ is the beam energy in the $\EE$ C.M.\ frame and $M_{\Xi_{c}^0}$ is the invariant mass of $\Xi_{c}^0$ candidates. All these selection criteria are optimized using the same method that was used for the energy of photon from the $\Sigma^0$ decay~\cite{input_Br, input_Br2}.

\section{\boldmath Branching fractions of $\Xi_c^0 \to \Lambda K_S^0$,
$\Xi_c^0 \to \Sigma^0 K_S^0$, and $\Xi_c^0 \to \Sigma^+ K^-$ decays}

\begin{figure}[htbp]
	\begin{center}
		\includegraphics[width=8cm]{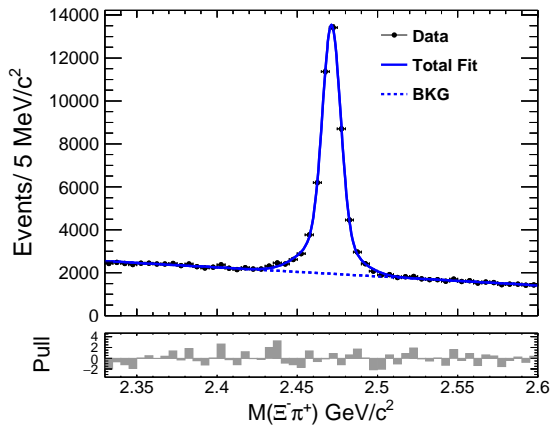}
		\caption{The invariant mass distribution of $\Xi^- \pip$ from data. The points with error bars
		represent the data, the blue solid curve shows the best-fit result, and the blue dashed curve
		represents the fitted background.}\label{fig2}
	\end{center}
\end{figure}

For the reference channel $\Xi_c^0 \to \Xi^-(\to \Lambda \pim)\pip$, except for the scaled momentum
$x_p$ and the $\Lambda$ selection criteria, all other selection criteria are similar to those
used in Ref.~\cite{semi_Xic}. The required $x_p$ value and the $\Lambda$ selection
criteria of the reference channel are the same as those of the signal channels.
Figure~\ref{fig2} shows the invariant mass distribution of $\Xi^- \pi^+$ with $x_p > 0.55$
from data, together with the results of an unbinned extended \mbox{maximum-likelihood} fit. In the fit,
the signal shape of $\Xi_c^0$ candidates is parameterized by a \mbox{double-Gaussian} function with different mean values, and
the background shape is described by a first-order polynomial. The parameters of signal
and background shapes are free. The fit result is displayed in Fig.~\ref{fig2} along with the pull
$(N_{\rm data}-N_{\rm fit})/\sigma_{\rm data}$ distribution, where $\sigma_{\rm data}$ is the uncertainty
on $N_{\rm data}$, and the fitted signal yield of $\Xi_c^0 \to \Xi^- \pip$ decay in data is $40539 \pm 315$.

\begin{figure*}[htbp]
	\begin{center}
		\includegraphics[width=5.9cm]{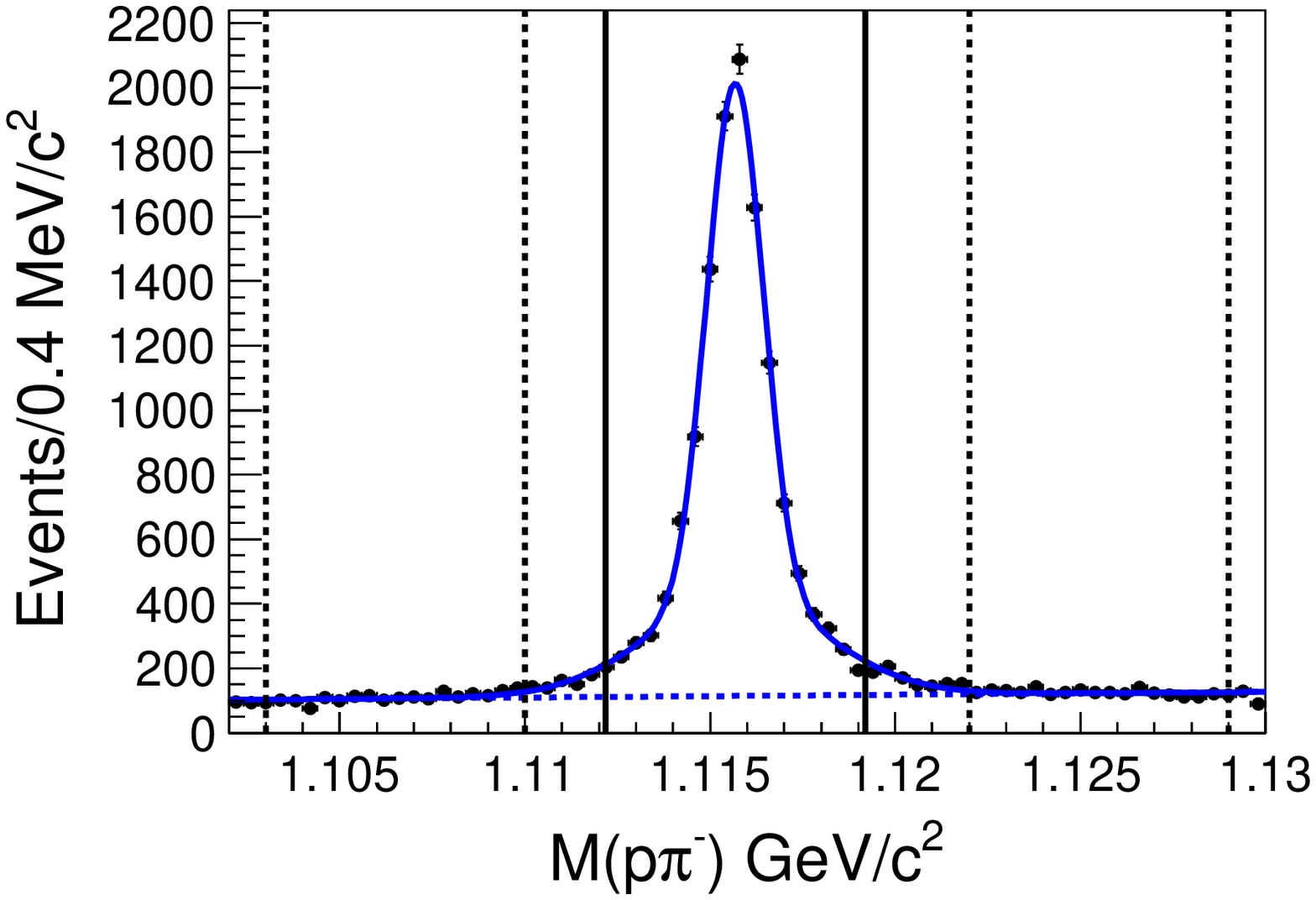}
		\includegraphics[width=5.9cm]{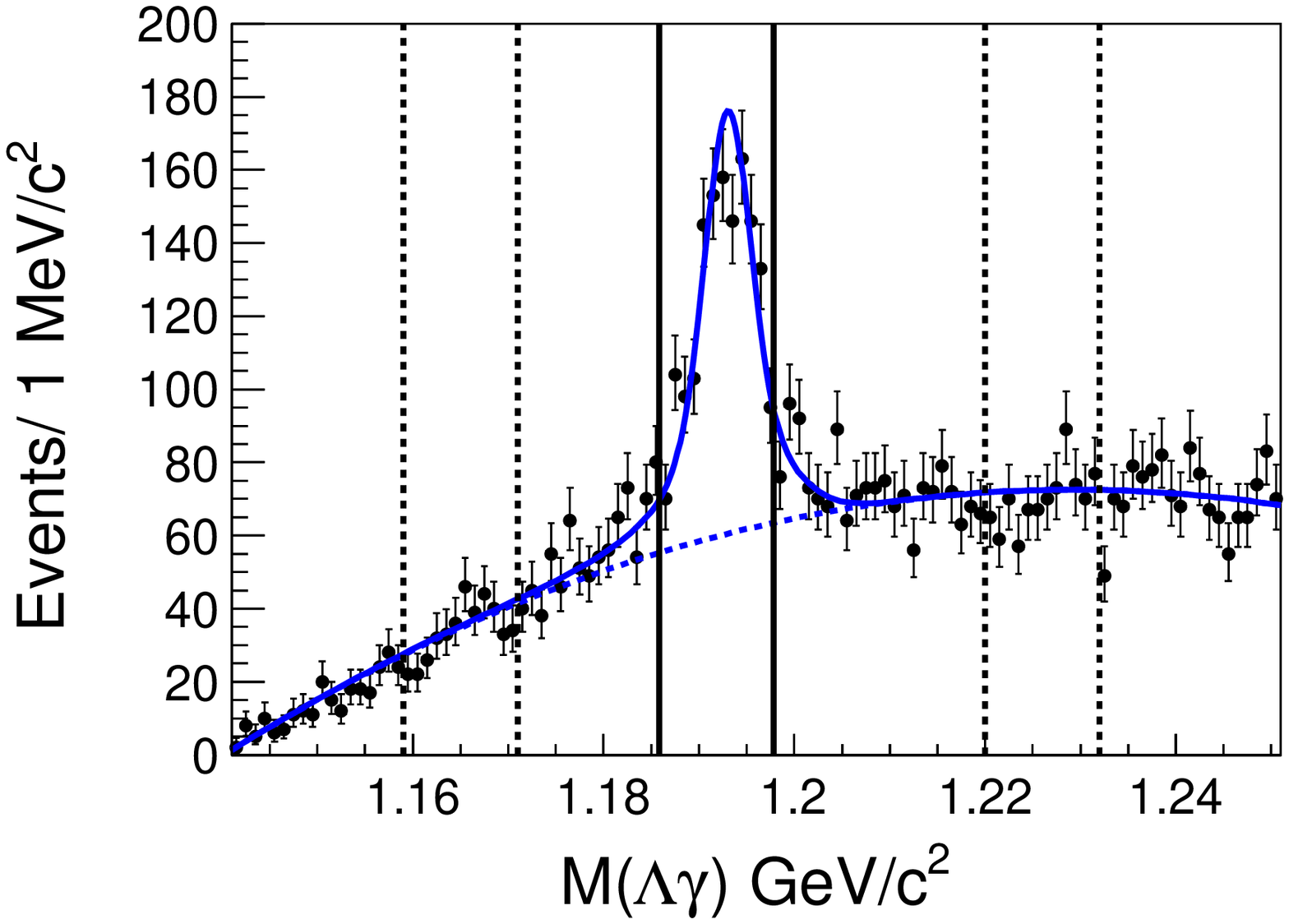}
		\includegraphics[width=5.9cm]{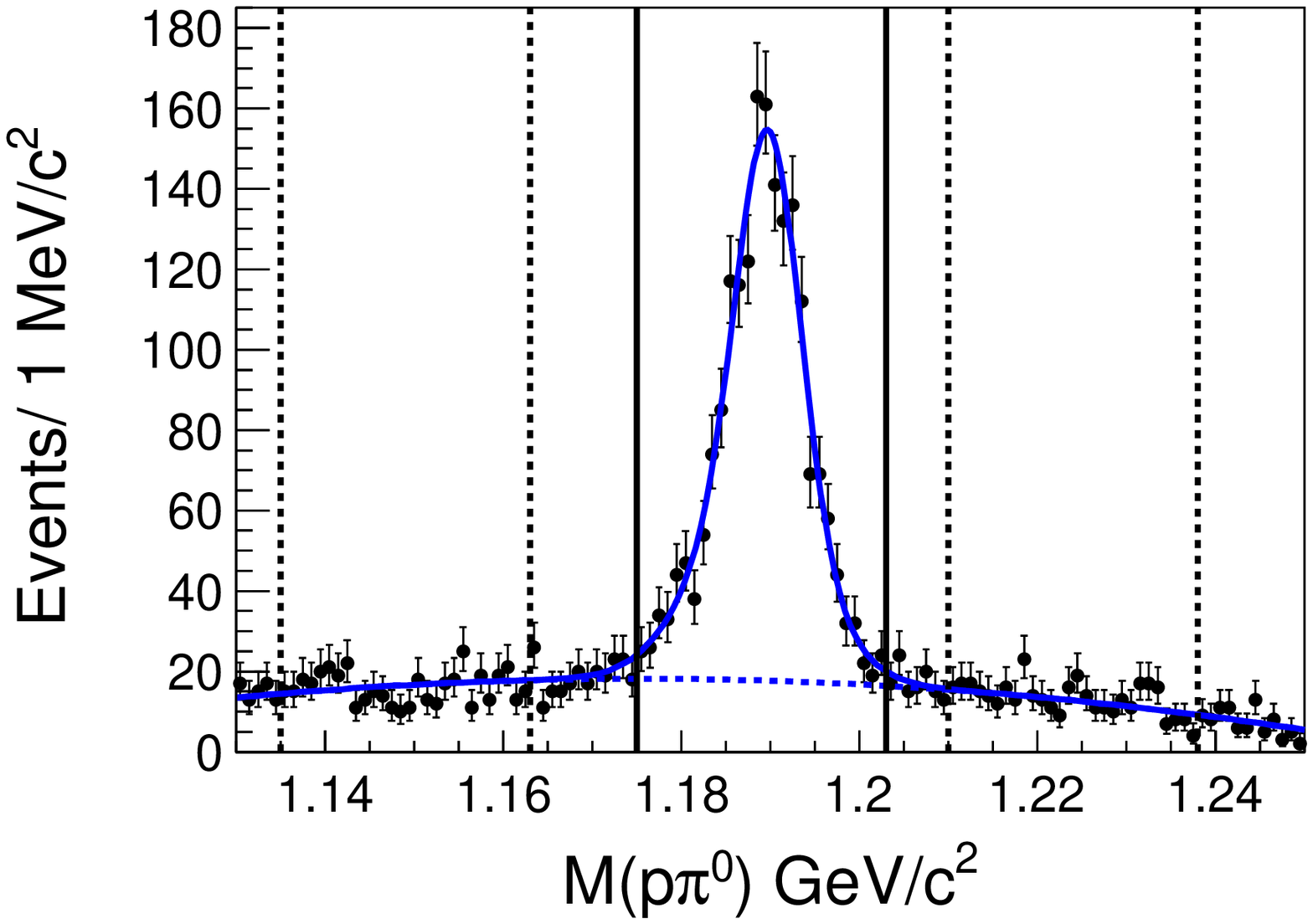}
		\put(-468,95){\bf (a)} \put(-300,95){\bf (b)}  \put(-125,95){\bf (c)}
		\caption{The invariant mass distributions of (a) $p\pi^-$, (b) $\Lambda\gamma$, and (c)
		$p\pi^0$ candidates from the decays $\Xi_c^0 \to \Lambda K_S^0$, $\Xi_c^0 \to \Sigma^0 K_S^0$, and
		$\Xi_c^0 \to \Sigma^+ K^-$ in the $\Xi_c^0$ signal region in data.
		The points with error bars represent the data, the blue solid curves show the best-fit results,
		and the blue dashed curves are the fitted backgrounds. The vertical solid lines represent the
		required signal regions, and the vertical dashed lines show the defined sidebands.}\label{fig3}
	\end{center}
\end{figure*}

After applying the aforementioned event selection criteria, the invariant mass
distributions of $p\pim$, $\Lambda\gamma$, and $p \pi^0$ from the decays
$\Xi_c^0 \to \Lambda K_S^0$, $\Xi_c^0 \to \Sigma^0 K_S^0$, and
$\Xi_c^0 \to \Sigma^+ K^-$ in data are shown
in Figs.~\ref{fig3}(a)$-$\ref{fig3}(c), together with the results of unbinned extended \mbox{maximum-likelihood} fits described below. There are significant $\Lambda$, $\Sigma^0$, and $\Sigma^+$ signals
observed in the $\Xi_c^0$ signal region, defined as a window of $\pm 20\mathrm{~MeV}/c^2$
around the $\Xi_c^0$ nominal mass~\cite{PDG} ($\sim2.5\sigma$).
In the fits, the signal shapes of $\Lambda$
and $\Sigma^+$ candidates are described by \mbox{double-Gaussian} functions with different mean values,
and the signal shape of $\Sigma^0$ is described by a \mbox{Crystal-Ball} function~\cite{CB}. The backgrounds are
parametrized by a first-order polynomial function for the $p\pi^-$ mass spectrum,
and second-order polynomial functions for the $\Lambda \gamma$ and $p\pi^0$ mass spectra.
The blue solid curves show the best-fit results, and the blue dashed curves represent the fitted backgrounds.
The reduced $\chi^2$ values of the fits are $\chi^2/\rm ndf = 1.20$, $1.23$, and $0.73$ for
$M(p\pim)$, $M(\Lambda\gamma)$, and $M(p \pi^0)$ distributions, respectively, where
$\rm ndf = 63$, $106$, and $112$ are the corresponding numbers of degrees of freedom.
The ratios of mass resolutions of $\Lambda$, $\Sigma^0$, and $\Sigma^+$ candidates
between the MC simulations and data are found to be $\sigma_{\rm MC}/\sigma_{\rm data} = 93\%$, $90\%$, and $92\%$,
respectively. The signal regions of $\Lambda$, $\Sigma^0$, and $\Sigma^+$ candidates
are defined as $\lvert M(p\pim ) - m(\Lambda) \rvert < 3.5\mathrm{~MeV}/c^2$,
$-7\mathrm{~MeV}/c^2 < M(\Lambda\gamma) - m(\Sigma^0) < 5\mathrm{~MeV}/c^2$,
and $\lvert M(p\pi^0) - m(\Sigma^+)\rvert < 14\mathrm{~MeV}/c^2$ with corresponding
efficiencies of approximately $95\%$, $83\%$, and $98\%$, respectively.
Here, $m(i)$ denotes the nominal mass of particle $i$~\cite{PDG}.
The above required signal regions are optimized using the same
method that was used for the energy of the photon from the $\Sigma^0$ decay.
We define the $\Lambda$, $\Sigma^0$, and $\Sigma^+$ sideband regions
as $1.103\mathrm{~GeV}/c^2 < M(p\pim) < 1.110\mathrm{~GeV}/c^2$ or $1.122\mathrm{~GeV}/c^2 < M(p\pim ) < 1.129\mathrm{~GeV}/c^2$,
$1.159\mathrm{~GeV}/c^2 < M(\Lambda\gamma) < 1.171\mathrm{~GeV}/c^2$ or $1.220\mathrm{~GeV}/c^2 < M(\Lambda\gamma) < 1.232\mathrm{~GeV}/c^2$,
and $1.135\mathrm{~GeV}/c^2 < M(p\pi^0) < 1.163\mathrm{~GeV}/c^2$ or $1.210\mathrm{~GeV}/c^2 < M(p\pi^0) < 1.238\mathrm{~GeV}/c^2$, respectively,
which are twice as wide as the corresponding signal regions. The vertical solid
lines indicate the required $\Lambda$, $\Sigma^0$, and $\Sigma^+$ signal regions, and the vertical dashed
lines represent the defined $\Lambda$, $\Sigma^0$, and $\Sigma^+$ sideband regions.

\begin{figure*}[htbp]
	\begin{center}
		\includegraphics[width=5.9cm]{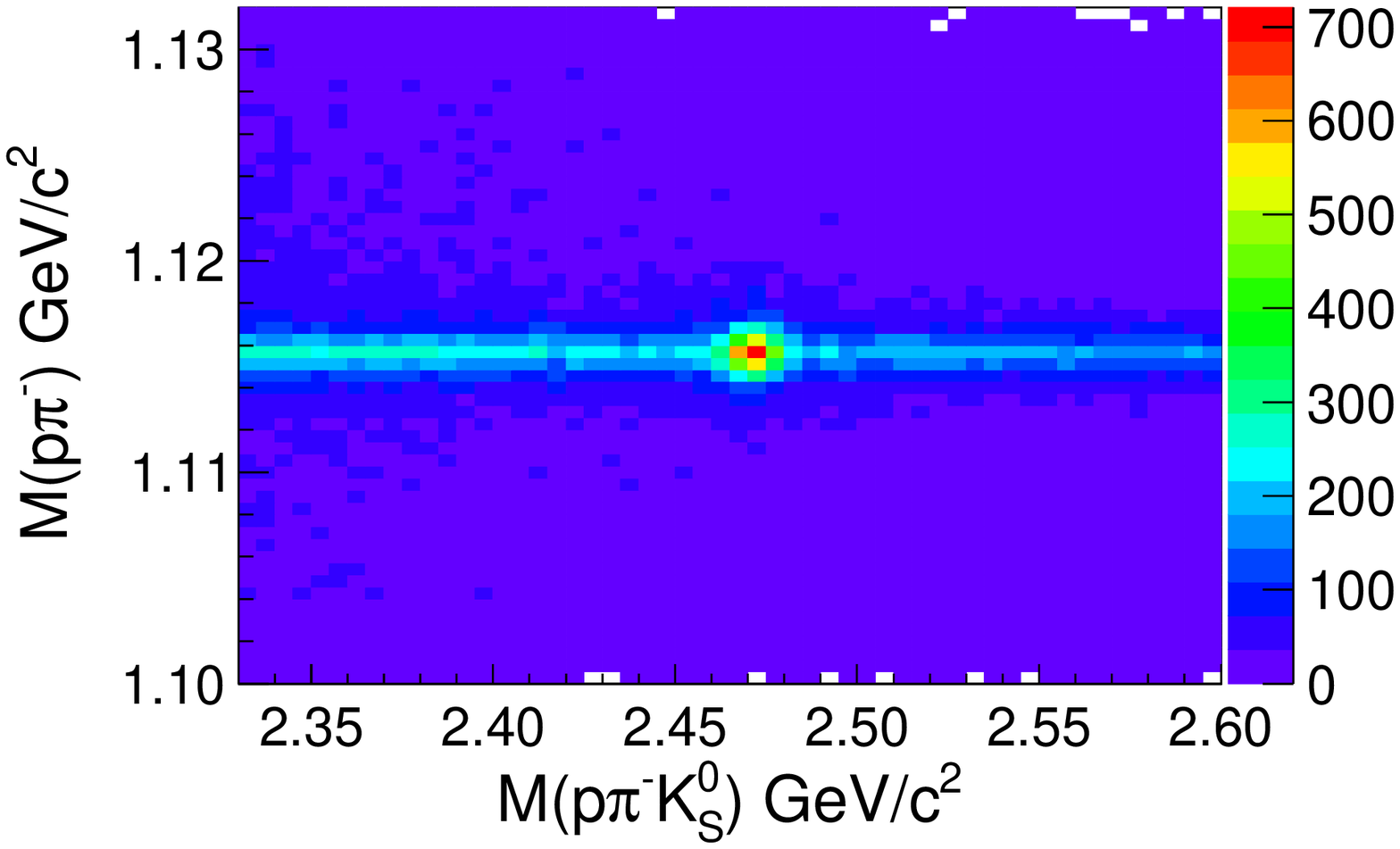}
		\includegraphics[width=5.9cm]{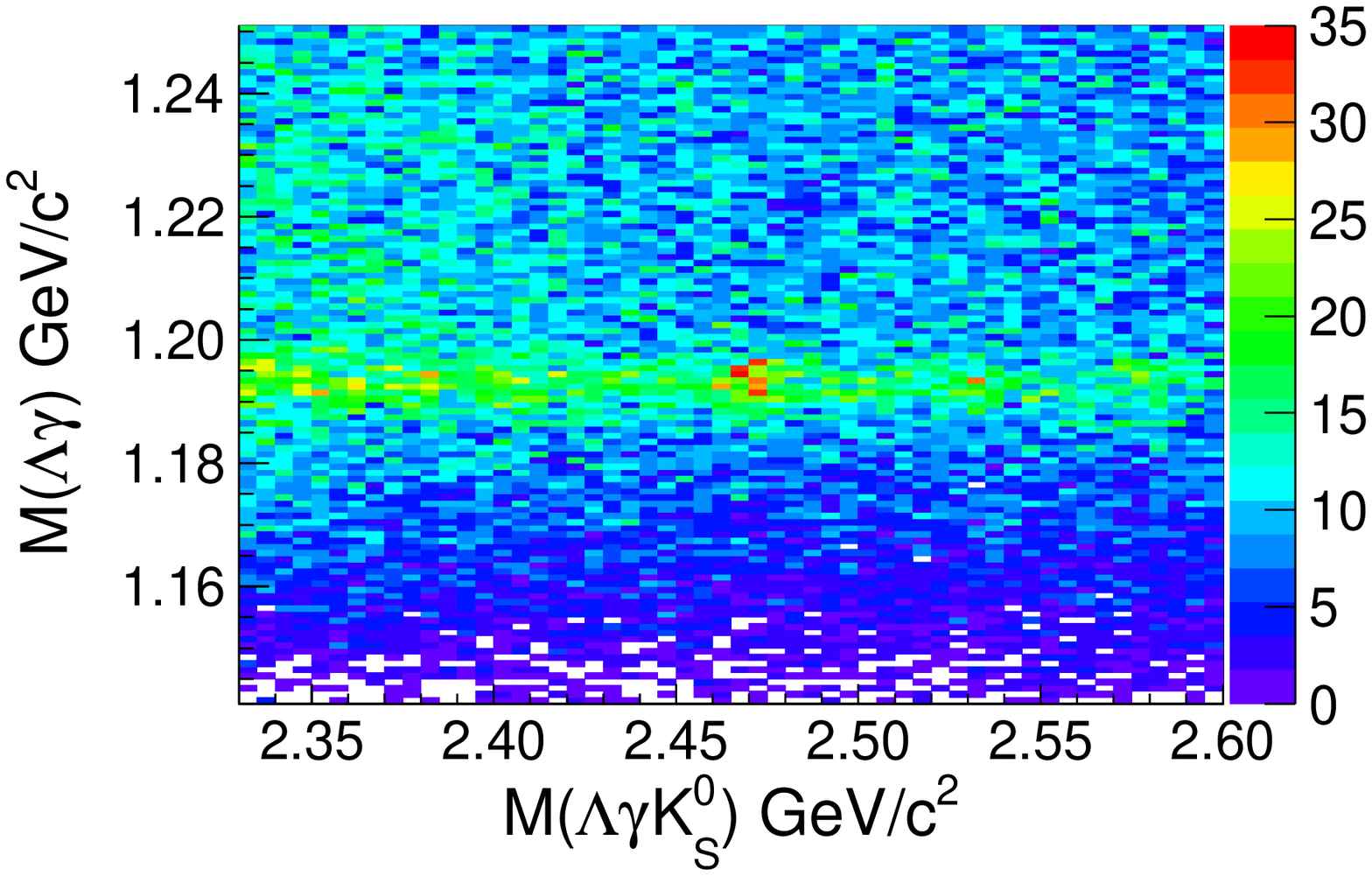}
		\includegraphics[width=5.9cm]{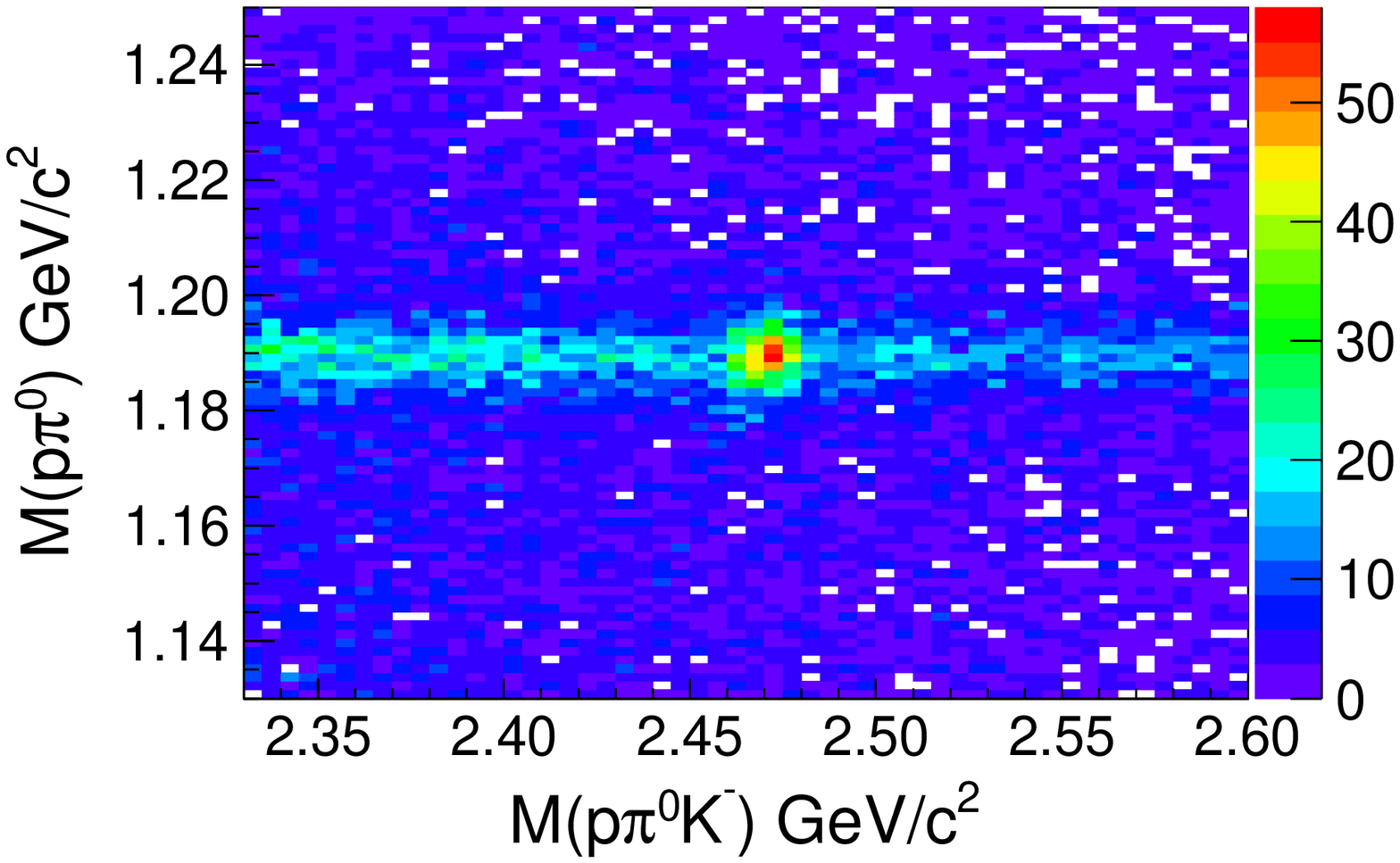}
		\put(-505,107){\bf (a)} \put(-335,107){\bf (b)}  \put(-165,107){\bf (c)}
		\caption{The scatter plots of (a) $M(p \pim)$ versus $M(p\pim K_S^0)$, (b) $M(\Lambda\gamma)$ versus $M(\Lambda\gamma K_S^0)$, and (c) $M(p \pi^0)$ versus $M(p\pi^0 K^-)$ from the selected $\Xi_c^0 \to \Lambda K_S^0$, $\Xi_c^0 \to \Sigma^0 K_S^0$, and $\Xi_c^0 \to \Sigma^+ K^-$ candidates in data.}\label{fig4}
	\end{center}
\end{figure*}

The scatter plots of $M(p\pi^-)$ versus $M(p\pim K_S^0)$, $M(\Lambda\gamma)$
versus $M(\Lambda \gamma K_S^0)$, and $M(p\pi^0)$ versus $M(p\pi^0 K^-)$ from data
are shown in Figs.~\ref{fig4}(a)$-$\ref{fig4}(c). From the plots,
significant $\Xi_c^0 \to \Lambda K_S^0$, $\Xi_c^0 \to \Sigma^0K_S^0$, and
$\Xi_c^0 \to \Sigma^+ K^-$ decays are observed.

\begin{figure*}[htbp]
	\begin{center}
		\includegraphics[width=5.9cm]{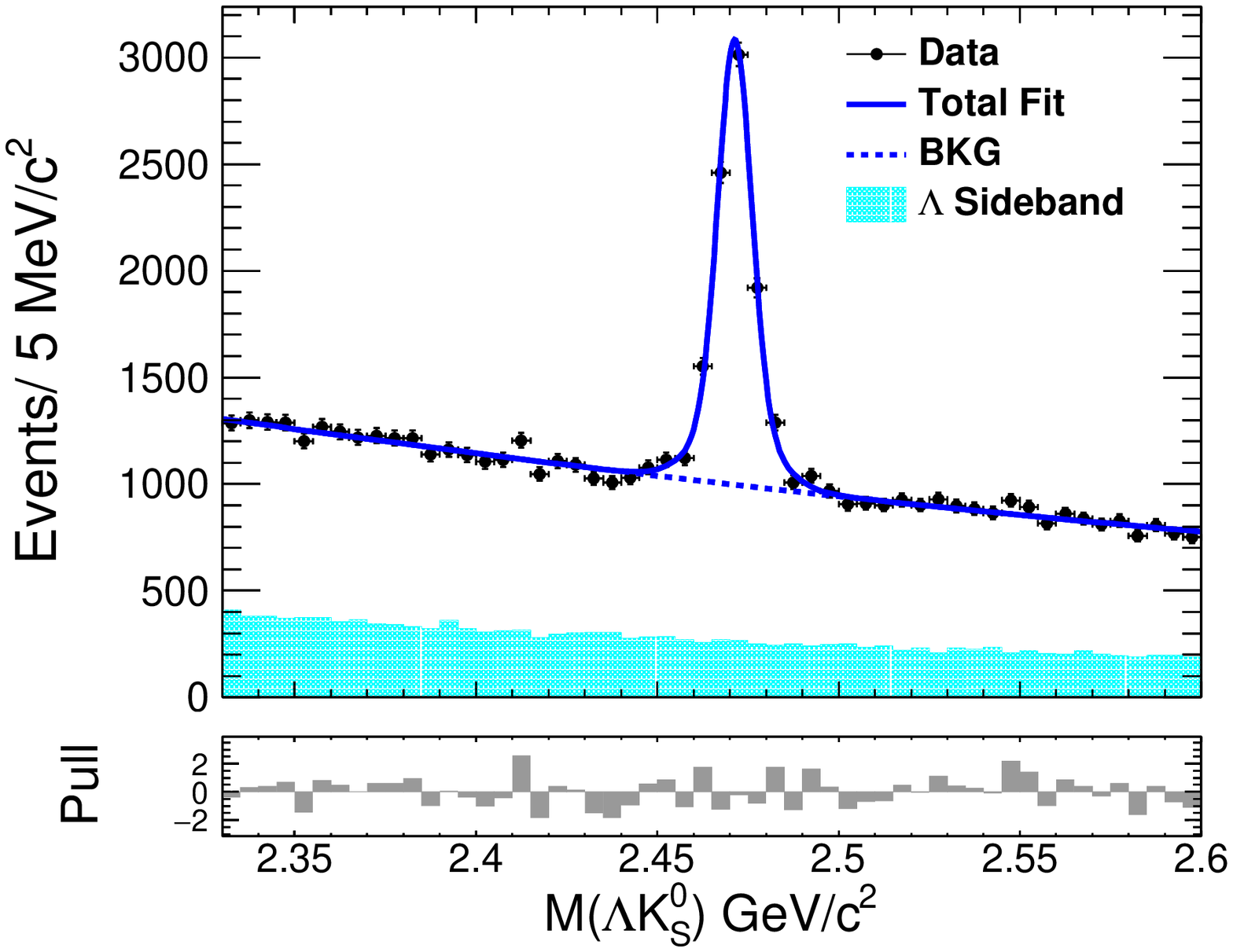}
		\includegraphics[width=5.9cm]{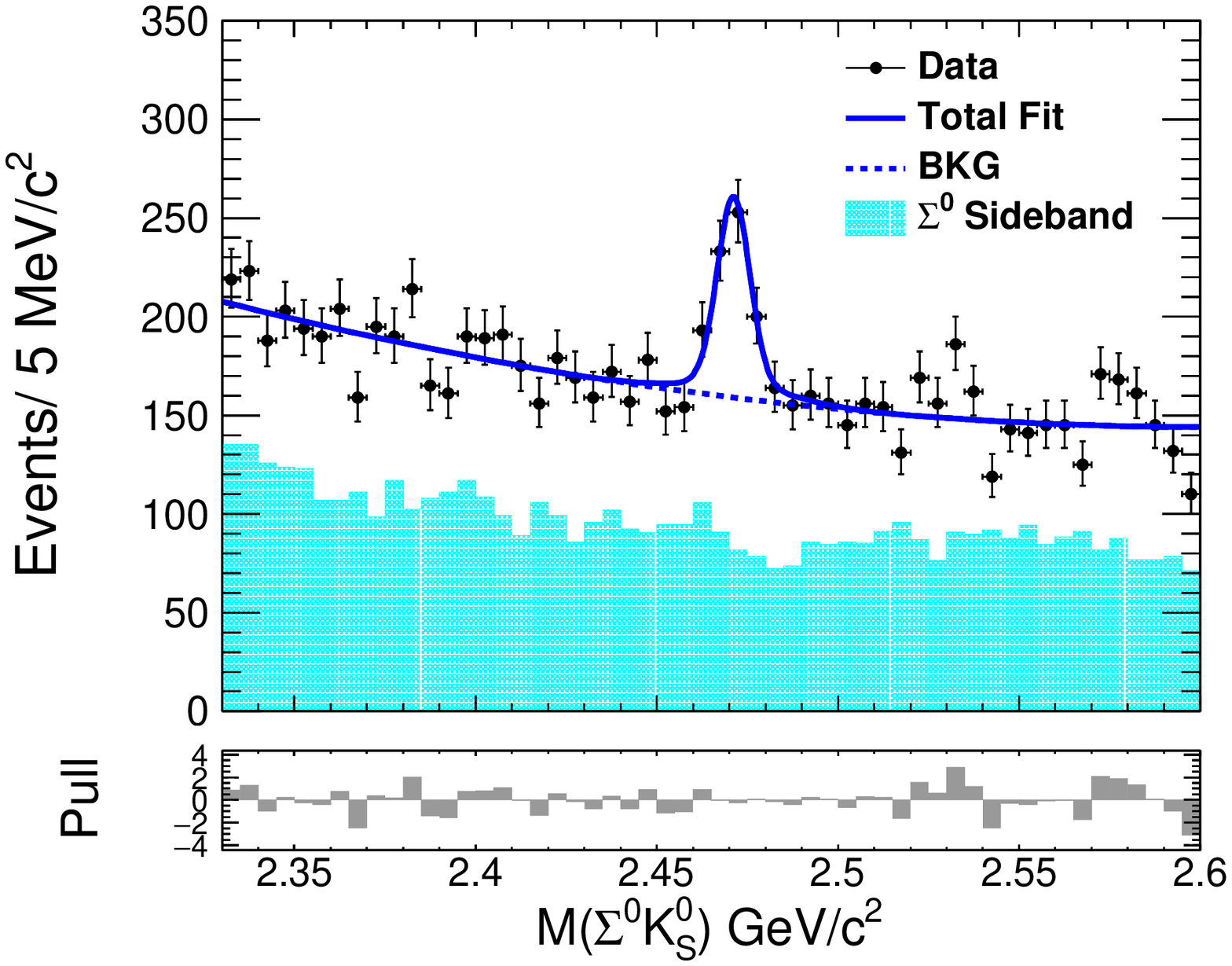}
		\includegraphics[width=5.9cm]{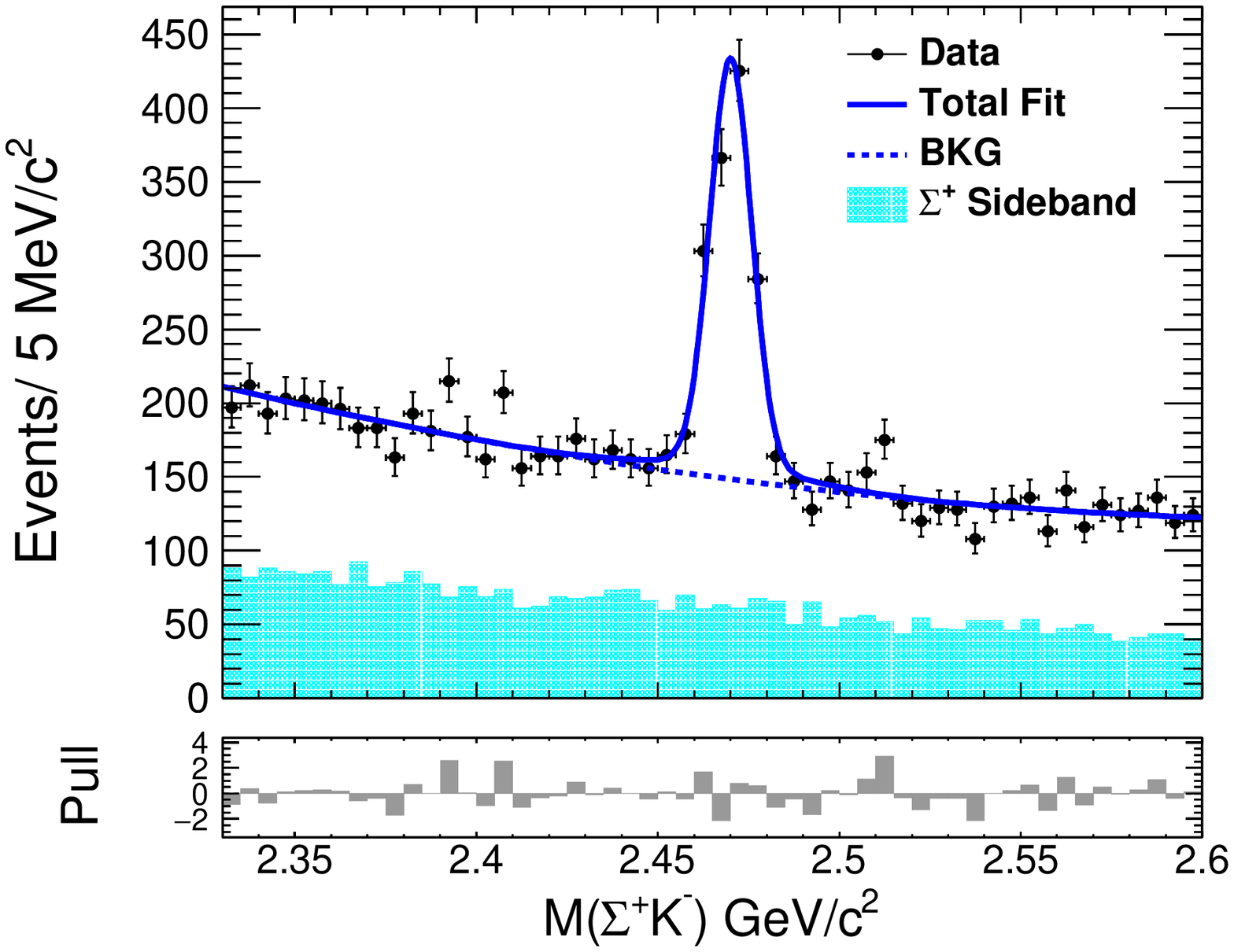}
		\put(-465,105){\bf (a)} \put(-295,105){\bf (b)}  \put(-125,105){\bf (c)}
		\caption{The invariant mass distributions of (a) $\Lambda K_S^0$, (b) $\Sigma^0K_S^0$,
		and (c) $\Sigma^+K^-$ from data. The points with error bars represent the data, the blue solid curves show the best-fit results, and
		the blue dashed curves show the fitted backgrounds. The cyan histograms represent
		events from the normalized $\Lambda$, $\Sigma^0$, and $\Sigma^+$ sidebands.}\label{fig5}
	\end{center}
\end{figure*}

Figure~\ref{fig5} shows the invariant mass spectra of $\Lambda K_S^0$, $\Sigma^0 K_S^0$,
and $\Sigma^+ K^-$ from data. The cyan shaded histograms indicate events
from the normalized $\Lambda$, $\Sigma^0$, and $\Sigma^+$ sidebands, respectively.
There are no evident peaking backgrounds found in the normalized sidebands or in the
inclusive MC samples. To extract the $\Xi_c^0$ signal yields from the two-body
decays $\Xi_c^0 \to \Lambda K_S^0$, $\Xi_c^0 \to \Sigma^0 K_S^0$, and
$\Xi_c^0 \to \Sigma^+ K^-$, we perform an unbinned extended maximum-likelihood
fit to each distribution. The signal shapes of $\Xi_c^0$ candidates are described by
double-Gaussian functions with different mean values, where the parameters
are floated for $\Xi_c^0 \to \Lambda K_S^0$ and $\Xi_c^0 \to \Sigma^+ K^-$
decays and are fixed to those obtained from the fit to the
corresponding simulated signal distribution for $\Xi_c^0 \to \Sigma^0 K_S^0$ decay.
The backgrounds are parametrized by second-order polynomial functions with free parameters.
The fit results are displayed in Fig.~\ref{fig5} along with the pull distributions, and
the corresponding reduced $\chi^2$ values of the fits are $\chi^2/\rm ndf = 1.12$, $1.44$, and $1.21$, respectively, where
$\rm ndf = 46$, $51$, and $46$ are the corresponding numbers of degrees of freedom.
The fitted mean values of $\Xi_c^0$ candidates in $\Xi_c^0 \to \Lambda K_S^0$ and
$\Xi_c^0 \to \Sigma^+ K^-$ decays are consistent with the $\Xi_c^0$ nominal mass~\cite{PDG},
and the fitted signal yields of $\Xi_c^0 \to \Lambda K_S^0$, $\Xi_c^0 \to \Sigma^0 K_S^0$,
and $\Xi_c^0 \to \Sigma^+ K^-$ decays in data are listed in Table~\ref{tab:summary}.
The statistical significances of $\Xi_c^0 \to \Lambda K_S^0$ and $\Xi_c^0 \to \Sigma^+ K^-$
decays are greater than $10\sigma$. The statistical significance of $\Xi_c^0 \to \Sigma^0 K_S^0$ decay
is $8.5\sigma$ calculated using $\sqrt{-2\ln(\mathcal{L}_{0}/\mathcal{L}_\text{max})}$, where $\mathcal{L}_{0}$ and
$\mathcal{L}_\text{max}$ are the maximized likelihoods without and with a signal component, respectively.

The branching fraction ratios of the decays $\Xi_c^0 \to \Lambda K_S^0/\Sigma^0 K_S^0/\Sigma^+ K^-$ relative
to that of $\Xi_c^0 \to \Xi^- \pi^+$ are calculated from the following formulae

\begin{align*}
	\frac{\BR(\Xi_c^0 \to \Lambda K_S^0)}{\BR(\Xi_c^0 \to \Xi^- \pip)}
	 = &~ \frac{N^{\rm obs}_{\Lambda K_S^0} \epsilon_{\Xi^- \pip}
	\BR(\Xi^- \to \Lambda \pim)}{N^{\rm obs}_{\Xi^- \pip} \epsilon_{\Lambda K_S^0}\BR(K_S^0\to\pip\pim)} \\
    = &~ 0.229\pm0.008(\rm stat.)\pm0.012(\rm syst.) \notag,
\end{align*}
\begin{align*}
	\frac{\BR(\Xi_c^0 \to \Sigma^0 K_S^0)}{\BR(\Xi_c^0 \to \Xi^- \pip)}
	 = &~ \frac{N^{\rm obs}_{\Sigma^0 K_S^0} \epsilon_{\Xi^- \pip}}{N^{\rm obs}_{\Xi^- \pip} \epsilon_{\Sigma^0 K_S^0}} \\ &
	\times \frac{\BR(\Xi^- \to \Lambda \pim)}{\BR(\Sigma^0 \to \Lambda \gamma) \BR(K_S^0\to\pip\pim)} \\
	= &~ 0.038\pm0.006(\rm stat.)\pm 0.004(\rm syst.)\notag,
\end{align*}
and
\begin{align*}
	\frac{\BR(\Xi_c^0 \to \Sigma^+ K^-)}{\BR(\Xi_c^0 \to \Xi^- \pip)}
    	= &~ \frac{N^{\rm obs}_{\Sigma^+ K^-}\epsilon_{\Xi^- \pip}}
	{N^{\rm obs}_{\Xi^- \pip}\epsilon_{\Sigma^+ K^-}} \\ &
    \times 	\frac{\BR(\Xi^- \to \Lambda \pi^-)\BR(\Lambda \to p \pi^-)}	{\BR(\Sigma^+ \to p \pi^0) \BR(\pi^0 \to \gamma \gamma)}
    \\ = &~ 0.123\pm0.007(\rm stat.)\pm 0.010(\rm syst.)\notag.
\end{align*}	
Here, $N^{\rm obs}_{\Lambda K_S^0}$, $N^{\rm obs}_{\Sigma^0 K_S^0}$, $N^{\rm obs}_{\Sigma^+ K^-}$, and $N^{\rm obs}_{\Xi^- \pip}$
are the fitted signal yields in decays $\Xi_c^0 \to \Lambda K_S^0$, $\Xi_c^0 \to \Sigma^0 K_S^0$, $\Xi_c^0 \to \Sigma^+ K^-$,
and $\Xi_c^0 \to \Xi^- \pi^+$, respectively; $\epsilon_{\Lambda K_S^0}$, $\epsilon_{\Sigma^0 K_S^0}$, $\epsilon_{\Sigma^+ K^-}$,
and $\epsilon_{\Xi^- \pip}$ are the corresponding reconstruction efficiencies, which are obtained from the signal MC simulations
and are listed in Table~\ref{tab:summary}. The efficiency correction factors of $95.5\%$ and $95.4\%$ from the required $\Sigma^0$ signal
region and PID of $\pi^+$ are included for $\epsilon_{\Sigma^0 K_S^0}$ and $\epsilon_{\Xi^- \pip}$, respectively, which are
discussed in Sec.V. Branching fractions $\BR(\Sigma^+ \to p \pi^0) = (51.57 \pm 0.30)\%$, $\BR(\pi^0 \to \gamma\gamma)$ = $(98.823\pm0.034)\%$,
and $\BR(\Lambda \to p \pim) = (63.9 \pm 0.5)\%$ are taken from Particle Data Group~\cite{PDG}.

\begin{table}[htbp]
	\caption{\label{tab:summary} Summary of the fitted signal yields $N^{\rm obs}$ and
		reconstruction efficiencies $\epsilon$. All the uncertainties here are
		statistical only.}
	\begin{tabular}{lr@{$\pm$}lr@{$\pm$}l}
		\hline\hline
		\multicolumn{1}{c}{Modes} & \multicolumn{2}{c}{$N^{\rm obs}$}  & \multicolumn{2}{c}{$\epsilon$(\%)}\\
		\hline
		$\Xi_c^0 \to \Lambda  K^0_S$  &  $5574$ & $180$  & $20.05$ & $0.08$ \\
		$\Xi_c^0 \to \Sigma^0 K^0_S$  &  $279$  & $41$   &  $6.03$ & $0.04$ \\	
		$\Xi_c^0 \to \Sigma^+ K^-$    &  $889$  & $50$   &  $5.15$ & $0.04$ \\	
		$\Xi_c^0 \to \Xi^-   \pip$    & $40539$ & $315$  & $23.24$ & $0.10$ \\	
		\hline\hline
	\end{tabular}
\end{table}

\section{\boldmath Systematic Uncertainties}
There are several sources of systematic uncertainties for the measurements of
branching fractions, including detection-efficiency-related uncertainties,
the branching fractions of intermediate states, as well as the overall fit uncertainties.
Note that the uncertainties from detection-efficiency-related sources
and the branching fractions of intermediate states partially cancel in the ratio to the
reference mode.

The detection-efficiency-related uncertainties include those from tracking efficiency,
PID efficiency, $K_S^0$ reconstruction efficiency, $\Lambda$ reconstruction efficiency,
photon reconstruction efficiency, $\pi^0$ reconstruction efficiency, and
the uncertainty related to the required $\Sigma^0$ signal region. Based on a study of
$D^{*+} \to \pi^+ D^0(\to K_S^0 \pi^+ \pi^-)$ decay, the tracking efficiency uncertainty
is evaluated to be $0.35\%$ per track. Using the $D^{*+} \to D^0\pip$, $D^0 \to K^{-}\pip$, and
$\Lambda \to p \pim$ control samples, the PID uncertainties are estimated to be $1.6\%$ per
kaon and $3.5\%$ per proton. The uncertainties associated with $K_S^0$, $\Lambda$, and $\pi^0$
reconstruction efficiencies are found to be $2.23\%$~\cite{KsUn}, $3.0\%$~\cite{LamUn},
and $2.25\%$~\cite{pi0Un}, respectively. The efficiency uncertainty in the photon reconstruction
is $2.0\%$ per photon, according to a study of radiative Bhabha events. For the
reference channel $\Xi_c^0 \to \Xi^-(\to\Lambda\pim)\pip$, the PID efficiency
uncertainties of $\pi^+$ from the $\Xi_c^0$ decay and $\pi^-$ from the $\Xi^-$ decay
are considered separately, because $\pi^+$ has a larger momentum. The PID efficiency ratio between the data
and MC simulation of $\pi^+$ is found to be $\epsilon_{\rm data}/\epsilon_{\rm MC} = (95.4 \pm 0.7)\%$, and then we take $95.4\%$ and $0.7\%$
as an efficiency correction factor and PID uncertainty for $\pip$; the PID efficiency ratio between the data
and MC simulation of $\pi^-$ is found to be $\epsilon_{\rm data}/\epsilon_{\rm MC} = (99.5 \pm 0.8)\%$, and $1.3\%$ is taken
as the PID uncertainty of $\pim$. We assume that $\Xi_c^0 \to \Lambda K_S^0/\Sigma^0 K_S^0/\Sigma^+ K^-$ decays are isotropic in the rest frame of $\Xi_c^0$, and a phase space model is used to generate signal events.
For the $\Xi_c^0 \to \Sigma^0 K_S^0$ decay, the $M(\Sigma^0)$ resolution discrepancy between data and MC simulation brings an efficiency correction factor $95.5\%$ and systematic uncertainty $0.5\%$ because of the required $\Sigma^0$ signal region.
For the $\Xi_c^0 \to \Lambda K_S^0$ and $\Xi_c^0 \to \Sigma^+ K^-$ decays, the uncertainties of the required $\Lambda$
and $\Sigma^+$ signal regions are less than $1\%$.
For the measurements of $\BR(\Xi_c^0\to\Lambda K_S^0)$ and $\BR(\Xi_c^0\to\Sigma^0 K_S^0)$, the uncertainties from tracking and $\Lambda$ reconstruction efficiencies mostly cancel by the reference channel. Assuming these uncertainties are independent and adding them in quadrature, the final detection-efficiency-related uncertainties are obtained, as listed in Table~\ref{tab:errsum}.

For the measurements of $\BR(\Xi_c^0 \to \Lambda K_S^0)$ and $\BR(\Xi_c^0 \to \Sigma^0 K_S^0)$,
the uncertainties from $\BR(\Xi^- \to \Lambda \pi^-)$ and $\BR(K_S^0 \to \pip \pim)$ are
$0.035\%$ and $0.072\%$~\cite{PDG}, which are small and neglected. For the measurement of
$\BR(\Xi_c^0 \to \Sigma^+ K^-)$, the uncertainties from $\BR(\Sigma^+ \to p\pi^0)$
and $\BR(\Lambda \to p \pim)$ are $0.6\%$ and $0.8\%$~\cite{PDG}, which are added in quadrature as
the total uncertainty from branching fractions of intermediate states.

The systematic uncertainties associated with the background shape, fit range,
and mass resolution are considered as follows. The order of the background
polynomial is changed from second to first or third, and the average deviation
compared to the nominal fit result is taken as the systematic uncertainty related to the background shape,
which are $3.26\%$, $9.11\%$, and $5.20\%$ for $\Xi_c^0 \to \Lambda K_S^0$, $\Xi_c^0 \to \Sigma^0 K_S^0$,
and $\Xi_c^0 \to \Sigma^+ K^-$ decays, respectively. The fit range is changed by $\pm 20\mathrm{~MeV}/c^2$, and the average deviation
compared to the nominal fit result is taken as the systematic uncertainty related to the fit range,
which are $1.67\%$, $2.14\%$, and $2.31\%$ for $\Xi_c^0 \to \Lambda K_S^0$, $\Xi_c^0 \to \Sigma^0 K_S^0$,
and $\Xi_c^0 \to \Sigma^+ K^-$ decays, respectively. For $\Xi_c^0 \to \Sigma^0 K_S^0$ decay, the signal shape of $\Xi_c^0$
is replaced by a Gaussian function with a free resolution convolved with the fixed signal shape from signal MC simulation:
the difference in the number of signal events, $4.32\%$, is taken as the systematic uncertainty related to the
mass resolution. The fit uncertainty of the reference mode is estimated using the same method as
was used for the signal modes, and the uncertainties associated with the background shape and
fit range are determined to be $1.54\%$ and $0.57\%$, respectively.
For each mode, all the above uncertainties are summed in quadrature
to obtain the total systematic uncertainty due to the fit.
Finally, the fit uncertainties of signal and reference modes are added in quadrature to give the total fit uncertainty
for each signal mode.

Assuming all the sources are independent and adding them in quadrature,
the total systematic uncertainties are obtained. All the systematical
uncertainties are summarized in Table~\ref{tab:errsum}.

\begin{table}[htbp]
	\caption{\label{tab:errsum}  Relative systematic uncertainties (\%) for the measurements of branching fractions of $\Xi_c^0 \to \Lambda K_S^0/\Sigma^0 K_S^0/\Sigma^+ K^-$. The uncertainty of $22.4\%$ on $\BR(\Xi_{c}^{0} \to \Xi^- \pi^{+})$~\cite{PDG} is treated as an independent systematic uncertainty.}
	\begin{tabular}{lr@{.}lr@{.}lc}
		\hline\hline
		\multicolumn{1}{c}{Sources} & \multicolumn{2}{c}{$\Xi_c^0 \to \Lambda K_S^0$} & \multicolumn{2}{c}{$\Xi_c^0 \to \Sigma^0  K_S^0$} & \multicolumn{1}{c}{$\Xi_c^0 \to \Sigma^+  K^-$}  \\
		\hline
		Detection efficiency & $3$     & $1$  &  $3$     & $7$  &  $5.9$   \\
		Branching fraction   & $< 0$   & $1$  &  $< 0$   & $1$  &  $1.0$   \\	
     	Fit uncertainty      & $4$     & $0$  &  $10$    & $5$  &  $5.9$   \\	
     	\hline
     	Sum in quadrature    & $5$     & $1$  & $ 11$    & $1$  &  $8.4$   \\
		\hline\hline
	\end{tabular}
\end{table}

\section{\boldmath Summary}
In summary, using the entire data sample of $980\mathrm{~fb}^{-1}$ integrated luminosity
collected with the Belle detector, we study $\Xi_c^0 \to \Lambda K_S^0$,
$\Xi_c^0 \to \Sigma^0 K_S^0$, and $\Xi_c^0 \to \Sigma^+ K^-$ decay modes. The ratios of the branching fractions of $\Xi_c^0 \to \Lambda  K_S^0$,
$\Xi_c^0 \to \Sigma^0 K_S^0$, and $\Xi_c^0 \to \Sigma^+ K^-$ relative to that of $\Xi_c^0 \to \Xi^- \pi^+$
are measured to be $0.229 \pm 0.008(\rm stat.) \pm 0.012(\rm syst.)$, $0.038 \pm 0.006(\rm stat.) \pm 0.004(\rm syst.)$,
and $0.123 \pm 0.007(\rm stat.) \pm 0.010(\rm syst.)$, respectively. The measured branching fraction ratio $\BR(\Xi_c^0 \to \Lambda  K_S^0)/\BR(\Xi_c^0 \to \Xi^- \pip)$ is consistent with the previously measured value of $0.21\pm0.02(\rm stat.) \pm 0.02(\rm syst.)$~\cite{Belle:2004tfm} with much improved precision and supersedes the previous result. Taking $\BR(\Xi_c^0 \to \Xi^- \pi^+) = (1.43 \pm 0.32)\%$~\cite{PDG}, the absolute branching fractions are determined to be

\begin{displaymath}
	\begin{aligned}
    \BR(\Xi_c^0 \to \Lambda K_S^0) = (3.27\pm0.11\pm0.17\pm0.73)\times10^{-3}, \notag
   \end{aligned}
\end{displaymath}
\begin{displaymath}
	\begin{aligned}
	\BR(\Xi_c^0 \to \Sigma^0 K_S^0) = (0.54\pm0.09\pm0.06\pm0.12)\times10^{-3}, \notag
   \end{aligned}
\end{displaymath}
\begin{displaymath}
	\begin{aligned}
   \BR(\Xi_c^0 \to \Sigma^+ K^-) = (1.76\pm0.10\pm0.14\pm 0.39)\times10^{-3}, \notag
   \end{aligned}
\end{displaymath}
where the uncertainties are statistical, systematic, and from $\BR(\Xi_c^0 \to \Xi^- \pip)$, respectively.
The branching fractions of $\Xi_c^0 \to \Sigma^0K_S^0$ and $\Xi_c^0 \to \Sigma^+ K^-$
decays, which are measured for the first time, are of the same order of magnitude as the theoretical predictions in Refs.~\cite{Zou:2019kzq,Geng:2019xbo}, but
an order of magnitude smaller than the predicted values in Ref.~\cite{Zhao:2018mov}.
All these measured branching fractions are in the same order of magnitude as the theoretical
predictions~\cite{Zou:2019kzq,Geng:2019xbo,Zhao:2018mov}. The measured ratios of
the branching fractions among the three decay modes are consistent with the theoretical predictions based on $SU(3)_{F}$
flavor symmetry approaches within the theoretical uncertainties~\cite{Geng:2019xbo,Zhao:2018mov}, but contradict
those predicted by dynamical model calculations~\cite{Zou:2019kzq}.

We thank the KEKB group for the excellent operation of the
accelerator; the KEK cryogenics group for the efficient
operation of the solenoid; and the KEK computer group, and the Pacific Northwest National
Laboratory (PNNL) Environmental Molecular Sciences Laboratory (EMSL)
computing group for strong computing support; and the National
Institute of Informatics, and Science Information NETwork 5 (SINET5) for
valuable network support.  We acknowledge support from
the Ministry of Education, Culture, Sports, Science, and
Technology (MEXT) of Japan, the Japan Society for the
Promotion of Science (JSPS), and the Tau-Lepton Physics
Research Center of Nagoya University;
the Australian Research Council including grants
DP180102629, 
DP170102389, 
DP170102204, 
DP150103061, 
FT130100303; 
Austrian Science Fund (FWF);
the National Natural Science Foundation of China under Contracts
No.~11475187,  
No.~11521505,  
No.~11575017,  
No.~11675166,  
No.~11705209;  
No.~11761141009;
No.~11975076;
No.~12042509;
No.~12135005;
Key Research Program of Frontier Sciences, Chinese Academy of Sciences (CAS), Grant No.~QYZDJ-SSW-SLH011; 
the  CAS Center for Excellence in Particle Physics (CCEPP); 
the Ministry of Education, Youth and Sports of the Czech
Republic under Contract No.~LTT17020;
the Carl Zeiss Foundation, the Deutsche Forschungsgemeinschaft, the
Excellence Cluster Universe, and the VolkswagenStiftung;
the Department of Science and Technology of India;
the Istituto Nazionale di Fisica Nucleare of Italy;
National Research Foundation (NRF) of Korea Grant
Nos.~2016R1\-D1A1B\-01010135, 2016R1\-D1A1B\-02012900, 2018R1\-A2B\-3003643,
2018R1\-A6A1A\-06024970, 2018R1\-D1A1B\-07047294, 2019K1\-A3A7A\-09033840,
2019R1\-I1A3A\-01058933;
Radiation Science Research Institute, Foreign Large-size Research Facility Application Supporting project, the Global Science Experimental Data Hub Center of the Korea Institute of Science and Technology Information and KREONET/GLORIAD;
the Polish Ministry of Science and Higher Education and
the National Science Center;
the Ministry of Science and Higher Education of the Russian Federation, Agreement 14.W03.31.0026; 
the Slovenian Research Agency;
Ikerbasque, Basque Foundation for Science, Spain;
the Swiss National Science Foundation;
the Ministry of Education and the Ministry of Science and Technology of Taiwan;
and the United States Department of Energy and the National Science Foundation.

\end{document}